\documentclass[showpacs,preprintnumbers,amsmath,amssymb,twocolumn]{revtex4}
\usepackage{graphicx}
\usepackage{dcolumn}
\usepackage{bm}
\usepackage{amssymb}
\usepackage{array}
\usepackage{longtable}

\begin{document}

\title{Trapping in scale-free networks with hierarchical organization of modularity}

\author{Zhongzhi Zhang$^{1,2}$}
\email{zhangzz@fudan.edu.cn}

\author{Yuan Lin$^{1,2}$}

\author{Shuyang Gao$^{1,2}$}

\author{Shuigeng Zhou$^{1,2}$}
\email{sgzhou@fudan.edu.cn}

\author{Jihong Guan$^{3}$}
\email{jhguan@tongji.edu.cn}

\author{Mo Li$^{4}$}

\affiliation {$^{1}$School of Computer Science, Fudan University,
Shanghai 200433, China}

\affiliation {$^{2}$Shanghai Key Lab of Intelligent Information
Processing, Fudan University, Shanghai 200433, China}

\affiliation{$^{3}$Department of Computer Science and Technology,
Tongji University, 4800 Cao'an Road, Shanghai 201804, China}

\affiliation {$^{4}$Software School, Fudan University, Shanghai
200433, China}

\begin{abstract}
A wide variety of real-life networks share two remarkable generic
topological properties: scale-free behavior and modular
organization, and it is natural and important to study how these two
features affect the dynamical processes taking place on such
networks. In this paper, we investigate a simple stochastic
process---trapping problem, a random walk with a perfect trap fixed
at a given location, performed on a family of hierarchical networks
that exhibit simultaneously striking scale-free and modular
structure. We focus on a particular case with the immobile trap
positioned at the hub node having the largest degree. Using a method
based on generating functions, we determine explicitly the mean
first-passage time (MFPT) for the trapping problem, which is the
mean of the node-to-trap first-passage time over the entire network.
The exact expression for the MFPT is calculated through the
recurrence relations derived from the special construction of the
hierarchical networks. The obtained rigorous formula corroborated by
extensive direct numerical calculations exhibits that the MFPT grows
algebraically with the network order. Concretely, the MFPT increases
as a power-law function of the number of nodes with the exponent
much less than 1. We demonstrate that the hierarchical networks
under consideration have more efficient structure for transport by
diffusion in contrast with other analytically soluble media
including some previously studied scale-free networks. We argue that
the scale-free and modular topologies are responsible for the high
efficiency of the trapping process on the hierarchical networks.
\end{abstract}

\pacs{05.40.Fb, 89.75.Hc, 05.60.Cd, 89.75.Da}


\date{\today}
\maketitle

\section{Introduction}

Complex networks are a powerful and versatile mathematical tool for
representing and modeling structure of complex
systems~\cite{AlBa02,DoMe02}, and their wide applications in
different areas have made them become a subject of a large volume of
research in the past decade~\cite{Ne03,BoLaMoChHw06}. Within the
general framework of complex networks, scientists can offer in
qualitative terms the detailed microscopic description of structural
properties and complexity of real-life systems. Extensive empirical
analysis on diverse real systems has unveiled that many, perhaps
most, real-world networks are simultaneously characterized by the
two most remarkable features: scale-free behavior~\cite{BaAl99} and
modular organization~\cite{GiNe02,RaSoMoOlBa02,RaBa03}. The
scale-free nature of a network means that its degree distribution
$P(k)$ follows a power law as $P(k) \sim k^{-\gamma}$ with the
degree distribution exponent in the range of $2< \gamma \leq 3$,
while the modular organization implies that the network is formed by
groups (modules) of nodes that have a significantly higher
interconnection density compared to the overall density of the whole
network. The important finding of these two fundamental natures has
led to the rising of research on some outstanding issues in the
field of complex networks such as exploring the generation
mechanisms for scale-free behavior~\cite{AlBa02,DoMe02}, detecting
and characterizing modular
structure~\cite{DaDuDiAr05,PaDeFaVi05,Ne06,Fo09}, and so on. On the
other hand, it has been shown that the two characteristics are
closely related to other structural properties such as average path
length~\cite{ChLu02,CoHa03} and clustering
coefficient~\cite{RaBa03}.

In principle, one of the main reasons for studying structural
properties of complex networks is to understand how the dynamical
processes are influenced by the underlying topological
structure~\cite{DoGoMe08}. Among a plethora of random processes,
random walks with wide range of distinct applications to many
science branches, have attracted a considerable amount of recent
attention within the physics
community~\cite{HaBe87,MeKl00,MeKl04,BuCa05,SoMaBl97,PaAm04,NoRi04,BeCoMoSuVo05,SoRebe05,Bobe05,GaSoHaMa07,BaBeWi08,BaCaPa08,KiCaHaAr08,ZhZhZhYiGu09,HaRo09}.
Particularly, trapping issue, an integral major theme of random
walks, is relevant to a variety of contexts, including target
research~\cite{JaBl01,Sh05}, photon-harvesting processes in
photosynthetic cells~\cite{WhGo99}, and characterizing similarities
between the elements of a database~\cite{FoPiReSa07} has led to an
increasing number of theoretical and practical investigations over
the last several decades. Numerous authors have made concerted
efforts to study trapping problem in different media, including
regular lattices~\cite{Mo69}, Sierpinski
fractals~\cite{KaBa02PRE,KaBa02IJBC}, \emph{T} fractal~\cite{Ag08},
small-world networks~\cite{CaAb08}, and scale-free
networks~\cite{ZhQiZhXiGu09,ZhZhXiChLiGu09,ZhGuXiQiZh09,AgBu09,ZhXiZhGaGu09},
as well as other
structures~\cite{CoBeTeVoKl07,BeMeTeVo08,CoTeVoBeKl08}. These
studies unclosed many unusual and exotic phenomena of trapping on
diverse graphs. However, the trapping process on scale-free networks
with modular structure remains less understood, in spite of the
facts that modularity plays an important role in shaping up
scale-free networks~\cite{GuSaAm07}, and that taking into account
the modular structure of scale-free networks leads to a better
understanding of how the underlying systems work~\cite{Ma07}.

In this paper, we study the classic trapping problem on a class of
hierarchical networks~\cite{RaSoMoOlBa02,RaBa03}, which is a random
walk problem with a single immobile trap positioned at a given site,
absorbing all walks visiting it. Here we focus on a particular case
with the trap located at the node with the highest degree. The
networks studied can capture simultaneously scale-free behavior and
modular structure. Moreover, the networks belong to a deterministic
growing type of networks, which have received much attention from
the scientific communities and have proved to be a useful
tool~\cite{BaRaVi01,DoGoMe02,JuKiKa02,CoOzPe00,ZhRoGo06,HiBe06,RoHaAv07,Hi07,ZhZhChGu07,ZhZhFaGuZh07,CoMi09}.
The deterministic nature of the hierarchical networks makes it
possible to investigate analytically the trapping process defined on
them. By applying the formalism~\cite{We1994,Re01} of generating
functions~\cite{Wi94} for random walks, we derive the rigorous
solution to the mean first-passage time (MFPT) that characterizes
the trapping process. The obtained exact result shows that the MFPT
scales algebraically with the number of network nodes. We also
compare the behavior of the trapping problem on the hierarchical
networks with those of other networks, and show that the
hierarchical networks can be helpful for enhancing the efficiency of
the trapping process.

\section{Modular scale-free networks}

\begin{figure}
\begin{center}
\includegraphics[width=0.85\linewidth,trim=50 280 130 245]{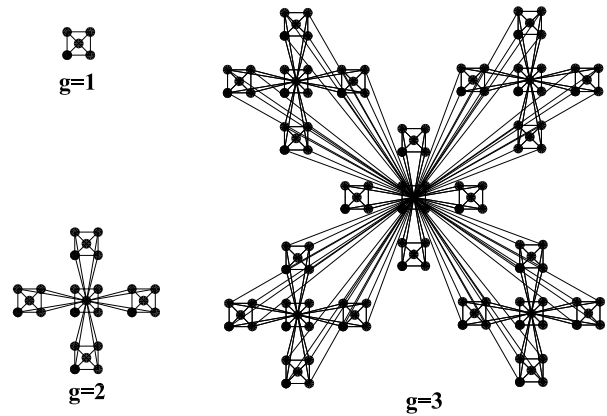}
\end{center}
\caption[kurzform]{\label{network} The iterative construction
process of a hierarchical network for the case of $M=5$. Notice that
the diagonal nodes are also connected
--- links not visible.}
\end{figure}

Let us introduce the model for the hierarchical scale-free networks
with a modular structure, which can be constructed in an iterative
way~\cite{RaSoMoOlBa02,RaBa03}. We denote by $H_{g}$ the network
model after $g$ ($g\geq 1$) iterations (number of generations).
Initially ($g=1$), the network consists of a central node, called
the hub (root) node, and $M-1$ peripheral (external) nodes with
$M\geq 3$. All these initial $M$ nodes are fully connected to each
other, forming a complete graph. At the second generation ($g=2$),
we generate $M-1$ copies of $H_{1}$ and connect the $M-1$ external
nodes of each replica to the root of the original $H_{1}$. The hub
of the original $H_{1}$ and the $(M-1)^2$ peripheral nodes in the
replicas become the hub and peripheral nodes of $H_{2}$,
respectively. Suppose one has $H_{g-1}$, the next generation network
$H_{g}$ can be obtained from $H_{g-1}$ by adding $M-1$ replicas of
$H_{g-1}$ with their external nodes being linked to the hub of the
original $H_{g-1}$ unit. In $H_{g}$, its hub is the hub of the
original $H_{g-1}$, and its external nodes are composed of all the
peripheral nodes of the $M-1$ copies of $H_{g-1}$. Repeating
indefinitely the replication and connection steps, we obtain the
hierarchical modular scale-free networks. Figure~\ref{network}
illustrates the construction process of a network for the particular
case of $M=5$, showing the first three iterations.

\begin{figure}
\begin{center}
\includegraphics[width=0.75\linewidth,trim=80 220 80 160]{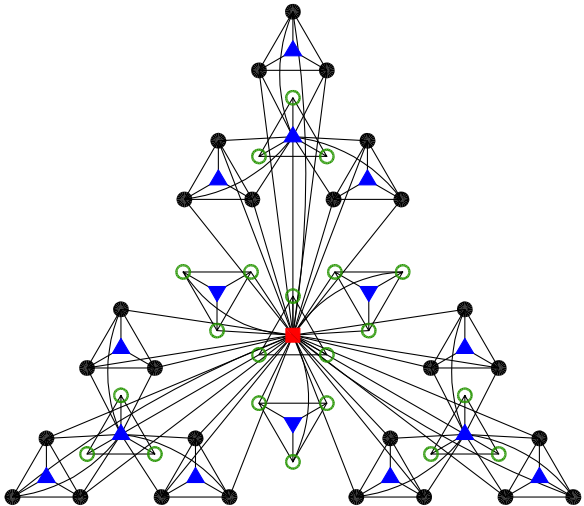}
\end{center}
\caption[kurzform]{(Color online) Classification of nodes in network
$H_3$ for the case of $M=4$. The filled circles, open circles, full
square, and triangles represent peripheral nodes, locally peripheral
nodes, hub node, and locally hub nodes, respectively.} \label{class}
\end{figure}

According to the network construction, one can see that $H_{g}$, the
network of $g$th generation, is characterized by two parameters $g$
and $M$, with the former being the number of generations, and the
latter representing the replication factor. In $H_{g}$, the number
of nodes, often called order of the network denoted as $N_g$, is
$N_g=M^{g}$. All these nodes can be classified into the following
four sets~\cite{No03,NoRi04a}: peripheral node set $\mathbb{P}$,
locally peripheral node set $\mathbb{P}_m$ ($1\leq m < g$), set
$\mathbb{H}$ only consisting of the hub node of $H_{g}$, and the
local hub set $\mathbb{H}_m$ ($1\leq m < g$); see Fig.~\ref{class}.
The cardinalities, defined as the number of nodes in a set, of the
four sets are
\begin{equation}\label{Car_K_p}
|\mathbb{P}| = (M-1)^{g},
\end{equation}
\begin{equation}\label{Car_K_lp}
|\mathbb{P}_m| = (M-1)^{m}M^{g-(m+1)},
\end{equation}
\begin{equation}\label{Car_K_h}
|\mathbb{H}| = 1,
\end{equation}
and
\begin{equation}\label{Car_K_lh}
|\mathbb{H}_m| = (M-1) M^{g-(m+1)},
\end{equation}
respectively. For $H_{g}$, all nodes belonging to the same set have
identical connectivity (i.e., degree), which are known exactly. For
example, the degree $K_h(g)$ of the hub node is the largest; it has
a value of
\begin{equation}\label{K_h}
K_h(g) = \sum_{g_i=1}^{g}(M-1)^{g_i}=\frac{M-1}{M-2}[(M-1)^g-1]\,.
\end{equation}
Any node in $\mathbb{P}$ has the degree
\begin{equation}\label{K_p}
K_p(g) = g+ M-2.
\end{equation}
Again, for instance, the degree of a node in $\mathbb{P}_m$ is
\begin{equation}\label{K_lp}
K_{p,m}(g) =m+ M-2\,,
\end{equation}
and an arbitrary node in $\mathbb{H}_m$ has a degree of
\begin{equation}\label{K_lh}
K_{h,m}(g) = \sum_{g_i=1}^{m}(M-1)^{g_i}=\frac{M-1}{M-2}
[(M-1)^m-1].
\end{equation}
Thus, the sum of degrees for all nodes in $H_g$ is
\begin{eqnarray}\label{calN}
&\quad& D_g\nonumber\\
&=& K_h(g)+\sum_{m=1}^{g-1}K_{h,m}(g)|\mathbb{H}_m|+K_p(g)|\mathbb{P}|+\sum_{m=1}^{g-1}K_{p,m}(g)|\mathbb{P}_m|\nonumber\\
&=&(3M-2)(M-1)M^{g-1} - 2 (M-1)^{g+1},
\end{eqnarray}
and the mean degree averaged over all nodes is
\begin{equation}\label{K_lh}
\langle k \rangle_g =
\frac{2D_g}{N_g}=2(M-1)\left[3-\frac{2}{M}-2\left(\frac{M-1}{M}\right)^g\right],
\end{equation}
which is approximately equal to $2(M-1)(3M-2)/M$ in the limit of
infinite $g$.

The hierarchical networks present some typical properties of real
systems in nature and society~\cite{No03}. They are scale free with
the degree distribution exponent $\gamma=1+\ln M / \ln (M-1)$. The
average path length, defined as the shortest distance averaged over
all pairs of nodes, scales logarithmically with the number of nodes.
In the large network order limit, the average clustering coefficient
tends to a large constant dependent on $M$. Thus, the whole family
of networks exhibits small-world behavior~\cite{WaSt98}. In
addition, the betweenness of nodes in the networks follows the same
power-law distribution $P_B \sim B^{-2}$ irrespective of $M$. In
particular, the networks show an obvious modular structure. All
these characteristics are not shared by other models. The peculiar
topological features make the networks unique within the category of
scale-free networks; it therefore is worthwhile to investigate
various dynamical processes running on them. In what follows we will
study the trapping problem on this class of modular networks to
uncover the influence of the particular topologies on the trapping
process.

\section{Formulation of the trapping problem}

\begin{figure}
\begin{center}
\includegraphics[width=0.8\linewidth,trim=100 240 100 190]{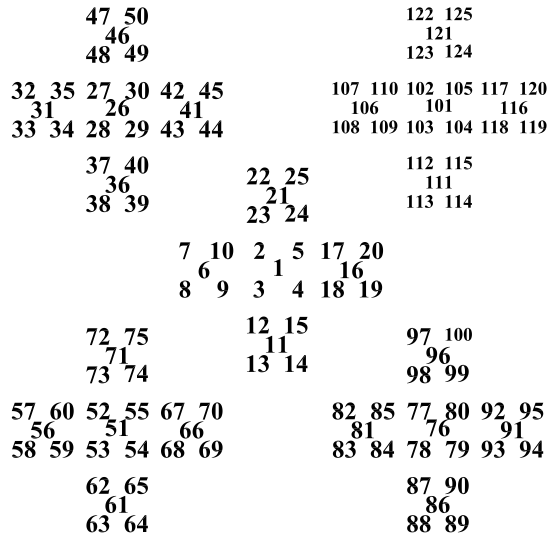}
\end{center}
\caption[kurzform]{\label{labeling} Labels of all nodes in ${H}_3$
for the case of $M=5$ corresponding to the $g=3$ case in
Fig.~\ref{network}.}
\end{figure}

In this section we formulate the trapping problem on the family of
hierarchical scale-free networks $H_g$, which is actually a simple
unbiased Markovian random walk of a particle in the presence of a
trap or a perfect absorber located on a given node. To facilitate
the description, we distinguish different nodes in $H_g$ by
assigning each of them a labeling in the following way. The hub node
in $H_g$ has label 1; the other $M-1$ peripheral nodes in $H_1$ are
labeled as 2, 3, and $M-1$, respectively. Assume that we have
labeled nodes in $H_{g-1}$ consecutively by 1, 2, and $M^{g-1}$; in
the next generation $g$, we keep the labels of nodes in the original
$H_{g-1}$ unchanged and label only the nodes belonging to the $M-1$
copies of $H_{g-1}$ by assigning to each node a different integer
from $M^{g-1}+1$ to $M^{g}$. In this way, every node in $H_g$ is
labeled by a unique integer from 1 to $N_g=M^{g}$; see
Fig.~\ref{labeling}.

For convenience, we continue to represent $H_g$ by its adjacency
matrix $\textbf{A}_g$ of order $N_g \times N_g$, whose $(i,j)$
element $a_{ij}$ is defined as follows: $a_{ij}=1$ if $i$ and $j$
are neighboring nodes and $a_{ij}=0$ otherwise. Then the degree,
$d_{i}(g)$, of node $i$ is given by $d_{i}(g)=\sum_{j}^{N_g}a_{ij}$,
the diagonal degree matrix $\textbf{Z}_g$ of $H_g$ is
$\textbf{Z}_g={\rm diag} (d_1(g), d_2(g),\ldots, d_i(g), \ldots,
d_{N_g}(g))$, and the normalized Laplacian matrix of $H_g$ is
provided by
$\textbf{L}_g=\textbf{I}_g-\textbf{Z}_g^{-1}\,\textbf{A}_g$, where
$\textbf{I}_g$ is the $N_g \times N_g$ identity matrix.

Before proceeding further, let us introduce the so-called
discrete-time random walk on $H_g$. At each time step, the particle
jumps from its current location to any of its nearest neighbors with
equal probability. According to this rule, at time $t$, a particle
located at a node $i$ will hop to one of its $d_i(g)$ neighbors, say
$u$, with the transition probability $a_{iu}/d_i(g)$. Suppose that
the particle starts off from node $i$ at $t=0$, then the jumping
probability $P_{ij}$ of going from $i$ to $j$ at time $t$ is
governed by the following master equation~\cite{NoRi04}:
\begin{equation}\label{master}
 P_{ij}(t+1)=\sum_{v=1}^{N_g} \frac{ a_{vj}}{ d_v(g)} \, P_{iv}(t).
\end{equation}

We next focus the trapping problem on $H_g$ with the trap fixed on
the hub node, i.e., node 1, represented as $i_T$. The particular
choice for the trap position allows to compute analytically the
MFPT, which will be discussed in detail in the following section.
Similar to the standard discrete-time random walks, during the
trapping process, in a single time step, the particle, starting from
any node except the trap $i_T$, jumps to any of its nearest
neighbors with the same probability. What we are concerned with is
the expected time that the particle spends, starting from a source
node before being trapped, which is in fact a random variable. Let
$X_i$ be the expected time, frequently called first-passage time
(FPT) or trapping time, for a walker, starting from node $i$, to
first arrive at the trap $i_T$. In order to determine $X_i$, we
define $F(X_i=t)$ to be the probability for the particle, starting
from point $i$, to first hit the trap after $t$ steps. Notice that
since the Markov chain~\cite{KeSn76} representing such a random walk
is ergodic, the particle will be eventually trapped independently of
the origin, implying that $\sum_{t=0}^{\infty}F(X_i=t)=1$ holds for
all $i$. It is easily known that the set of these interesting
quantities obeys the following recurrence relation:
\begin{equation} \label{MFPT01}
F(X_i=t)=\sum_{v=1}^{N_g} \frac{a_{iv}}{d_i(g)} F(X_v=t-1),
\end{equation}
where $i\neq i_T$.

Let $\tilde{F}_i(z)$ be the corresponding generating function of
quantity $F(X_i=t)$:
\begin{equation} \label{GenF01}
\tilde{F}_i(z)= \sum_{t=0}^{\infty} F(X_i = t)z^t,
\end{equation}
which encapsulates all the information contained in the discrete
probability distribution $F(X_i = t)$. For example, the expected
value $X_i$ is the first derivative of $\tilde{F}_i(z)$ evaluated at
$z=1$.

Let $\mathbf{\tilde{F}}(z)$ stand for the $(N_g-1)$-dimensional
vector $[\tilde{F}_2(z), \tilde{F}_3(z), \ldots,
\tilde{F}_{N_g}(z)]^{\top}$, where the superscript $\top$ represents
the transpose of the vector. According to Eqs.~(\ref{MFPT01}) and
~(\ref{GenF01}), we have
\begin{equation} \label{GenF02}
 \mathbf{\tilde{F}}(z)= z\mathbf{W}\,\mathbf{\tilde{F}}(z)\,,
\end{equation}
where $\mathbf{W}$ is a matrix with order $(N_g-1)\times (N_g-1)$
with entry $w_{ij}=a_{ij}/d_i(g)$. Differentiating the two sides of
Eq.~(\ref{GenF02}) with respect to $z$ and doing some simple algebra
operations, we have
\begin{equation}\label{GenF03}
        (\mathbf{I} - z\mathbf{W})\mathbf{\tilde{F}}'(z) - \mathbf{W}\mathbf{\tilde{F}}(z)=
        \mathbf{0},
\end{equation}
in which $\mathbf{I}$ is the identity matrix with order $(N_g-1)
\times (N_g-1)$; $\mathbf{0}$ is the $(N_g-1)$-dimensional zero
vector $(0, 0, \ldots, 0)^{\top}$. Setting $z=1$ in
Eq.~(\ref{GenF03}) leads to
\begin{equation}\label{MFPT02}
\mathbf{\tilde{F}}'(1)=(\mathbf{I}-\mathbf{W})^{-1}
\mathbf{W}\mathbf{\tilde{F}}(1)=(\mathbf{I}-\mathbf{W})^{-1}\mathbf{e},
\end{equation}
where $\mathbf{e} = (1, 1, \ldots, 1)^\top$ is the
$(N_g-1)$-dimensional unit vector.  Actually,
$(\mathbf{I}-\mathbf{W})^{-1}$ in Eq.~(\ref{MFPT02}) is the
fundamental matrix of the Markov chain representing the unbiased
random walk, and $\textbf{I}-\textbf{W}$ is a submatrix of the
normalized discrete Laplacian matrix $\textbf{L}_g$ of $H_g$, which
is obtained from $\textbf{L}_g$ by removing from $\textbf{L}_g$ the
first row and column corresponding to the trap.

From Eq.~(\ref{MFPT02}), the mean first-passage time, $\langle T
\rangle_g$, which is the average of $X_i$ over all initial nodes
distributed uniformly over nodes in $H_g$ other than the trap, is
given by
\begin{equation}\label{MFPT03}
 \langle T
\rangle_g=\frac{1}{N_g-1}\sum_{i=2}^{N_g}
X_i=\frac{1}{N_g-1}\sum_{i=2}^{N_g}\sum_{j=2}^{N_g}{l_{ij}},
\end{equation}
where $l_{ij}$ is the corresponding $(i,j)$ element of matrix
$(\mathbf{I}-\mathbf{W})^{-1}$, which is the mean time that the
particle spends at node $i$ starting from node $j$~\cite{BaKl98}.

\begin{figure}
\begin{center}
\includegraphics[width=0.7\linewidth,trim=95 100 120 40]{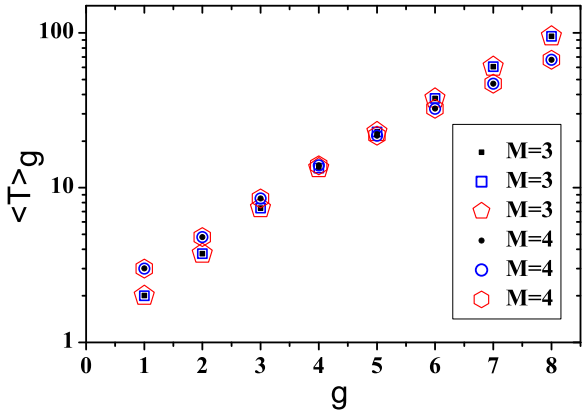}
\end{center}
\caption[kurzform]{\label{Time} (Color online) Mean first-passage
time $\langle T \rangle_g$ as a function of the iteration $g$ on a
semilogarithmic scale for two cases of $M=3$ and $M=4$. The filled
symbols are the data coming from genuine simulations of the trapping
process; the empty squares and circles represent the numerical
results obtained by direct calculation from Eq.~(\ref{MFPT03});
while the empty pentagons and hexagons correspond to the exact
values given by Eq.~(\ref{eq:MFPT09}).}
\end{figure}

Equation~(\ref{MFPT03}) shows that the problem of calculating MFPT
$\langle T \rangle_g$ is reduced to finding the sum of all elements
of matrix $(\mathbf{I}-\mathbf{W})^{-1}$. Since the order of
$(\mathbf{I}-\mathbf{W})$ is $(N_g-1)\times (N_g-1)$, where $N_g$
increases exponentially with $g$, for large $g$, inverting matrix
$(\mathbf{I}-\mathbf{W})$ is prohibitively time and memory
consuming, making it intractable to obtain $\langle T \rangle_g$
through direct calculation from Eq.~(\ref{MFPT03}); one can compute
directly the MFPT only for the first several generations (see
Fig.~\ref{Time}). Hence, an alternative method of computing MFPT
becomes necessary. In~\cite{BaLo06}, to allow for a drastic
reduction in computational cost, a scheme was proposed mapping the
original Markov process on another Markov process. Although the
method can bring down the computational efforts, it is an
approximate one. Fortunately, the special recursive construction of
the hierarchical networks allows to calculate analytically MFPT to
obtain an explicit solution for arbitrary generation $g$. In the
next section, we will provide the detailed process for the
derivation of MFPT using a method significantly different from that
applied
in~\cite{ZhQiZhXiGu09,ZhZhXiChLiGu09,ZhGuXiQiZh09,ZhXiZhGaGu09}.

\section{Closed-form solution to mean first-passage time}

Prior to deriving the general formula for MFPT, $\langle T
\rangle_g$, for the trapping issue on $H_g$, we first define some
related quantities. Let $P_g(t)$ denote the probability that, at the
generation $g$, the particle starting from any peripheral node in
$\mathbb{P}$ first arrives at the hub after $t$ jumps; and let
$Q_g(t)$ represent the probability that, the walker originating from
the hub to first reach any node belonging to $\mathbb{P}$ after $t$
steps. Then, the following fundamental relations can be established:
\begin{eqnarray}\label{MTT01}
P_g(t)=&\quad&\frac{\delta_{t,1}}{K_p(g)}
+\frac{M-2}{K_p(g)}P_g(t-1)\nonumber\\
&+&\frac{1}{K_p(g)}\sum_{m=1}^{g-1}\sum_{i=1}^{t-1}Q_m(i)P_g(t-1-i)
\end{eqnarray}
and
\begin{equation}\label{MTT02}
Q_g(t)=\frac{(M-1)^{g}}{K_h(g)}\delta_{t,1}+\sum_{m=1}^{g-1}\sum_{i=1}^{t-1}\frac{(M-1)^{m}}{K_h(g)}
P_m(i)Q_g(t-1-i),
\end{equation}
where $\delta_{t,1}$ is the Kronecker delta function that is defined
as follows: $\delta_{t,1}=1$ if $t$ is equal to 1, and
$\delta_{t,1}=0$ otherwise. Note that in Eqs.~(\ref{MTT01})
and~(\ref{MTT02}), the equivalence of nodes in the same set (e.g.,
$\mathbb{P}$ or $\mathbb{P}_m$) was used.

The three terms on the right-hand side (rhs) of Eq.~(\ref{MTT01})
can be elaborated as follows: the first term accounts for the
probability that the walker takes only one time step to first reach
the hub; the second term on the rhs explains the case that the
particle gets first to one of its $M-2$ neighbors belonging to
$\mathbb{P}$ in one time step, and then it takes more $t-1$ steps to
first arrive at the target node; the last term on the rhs describes
the probability of the process in which the walker first makes a
jump to a local hub node belonging to $\mathbb{P}_m$, then it takes
$i$ time steps, starting from the local hub, to hit one of the nodes
in $\mathbb{P}$, and continues to jump more $t-1-i$ steps to first
reach the hub.

Analogously, the two terms on the rhs of Eq.~(\ref{MTT02}) can be
understood based on the following two processes. The first term
explains the occurring probability of the process that the walker,
starting from the hub, only needs one time step to reach a
peripheral node in $\mathbb{P}$. The second term represents the
happening probability of such a process that the particle,
originating from the hub, first makes one jump to a local peripheral
node in $\mathbb{P}_m$, then makes $i$ jumps to the hub, and
proceeds to the destination (one of the nodes in $\mathbb{P}$),
taking more $t-1-i$ time steps.

Equations~(\ref{MTT01}) and~(\ref{MTT02}) provide the two basic
relations governing the trapping problem performing on $H_g$, from
which almost all subsequent results are derived from. As shown in
the preceding section, although we are concerned about only the
fundamental quantity (namely, MFPT), the direct calculations are
practically hard and intractable for large networks. Nevertheless,
the particular construction of the networks allows to overcome this
difficulty in virtue of the powerful mathematical technique of
generating functions~\cite{Wi94}, through which we can compute and
determine the MFPT $\langle F \rangle_g$ indirectly.

First, we define two generating functions, $\tilde{P}_g(x)$ and
$\tilde{Q}_g(x)$, for the probability distribution of first-passage
time described in Eqs.~(\ref{MTT01}) and~(\ref{MTT02}), which can be
written as
\begin{eqnarray}\label{eq:genP}
\tilde{P}_g(x)&=&\sum_{t=0}^{\infty}P_g(t)x^{t}\nonumber\\
&=&\frac{x}{K_p(g)}
+\frac{M-2}{K_p(g)}x\tilde{P}_g(x)+\frac{x}{K_p(g)}\sum_{m=1}^{g-1}\tilde{Q}_m(x)\tilde{P}_g(x)\nonumber\\
\end{eqnarray}
and
\begin{eqnarray}\label{eq:genQ}
\tilde{Q}_g(x)&=&\sum_{t=0}^{\infty}Q_g(t)x^{t}\nonumber\\&=&\frac{(M-1)^{g}}{K_h(g)}x+\frac{\tilde{Q}_g(x)}{K_h(g)}x\sum_{m=1}^{g-1}(M-1)^{m}\tilde{P}_m(x).\nonumber\\
\end{eqnarray}
After some algebraic operations, Eqs.~(\ref{eq:genP})
and~(\ref{eq:genQ}) can be recast, respectively, as
\begin{equation}\label{eq:genP1}
\tilde{P}_g(x)\left[\frac{K_p(g)}{x}-(M-2)-\sum_{m=1}^{g-1}\tilde{Q}_m(x)\right]=1
\end{equation}
and
\begin{equation}\label{eq:genQ1}
\tilde{Q}_g(x)\left[\frac{K_h(g)}{(M-1)^{g}}\frac{1}{x}-\sum_{m=1}^{g-1}(M-1)^{m-g}\tilde{P}_m(x)\right]=1.
\end{equation}

Let $T_g^P$ denote the first-passage time for a walker starting from
an arbitrary node in $\mathbb{P}$ to reach the hub for the first
time, which is in fact the number of steps for the walker
originating from any node in $\mathbb{P}$ to first visit the hub.
Let $T_g^H$ stand for the FPT needed for a particle initially
located at the hub to first hit any node in $\mathbb{P}$. Then,
according to the property of generating functions, the two
quantities $T_g^P$ and $T_g^H$ are given separately by
\begin{equation}\label{eq:MPTP}
T_g^P=\frac{d}{d x}\tilde{P}_g(x)\bigg |_{x=1}
\end{equation}
and
\begin{equation}\label{eq:MPTPQ}
T_g^H=\frac{d}{d x}\tilde{Q}_g(x)\bigg |_{x=1}.
\end{equation}

Differentiating, respectively, both sides of Eqs.~(\ref{eq:genP1})
and~(\ref{eq:genQ1}) with respect to $x$ and setting $x=1$, we
obtain the following two coupled relations:
\begin{equation}\label{eq:MPTP1}
T_g^P=g+M-2+\sum_{m=1}^{g-1}T_m^H
\end{equation}
and
\begin{equation}\label{eq:MPTPQ1}
T_g^H=\frac{K_h(g)}{(M-1)^{g}}+\frac{1}{(M-1)^{g}}\sum_{m=1}^{g-1}(M-1)^{m}T_{m}^{P}.
\end{equation}
From the above two coupled equations, it is not difficult to have
\begin{equation}\label{eq:MPTP2}
T_{g+1}^P-T_g^P=1+T_g^H
\end{equation}
and
\begin{equation}\label{eq:MPTPQ2}
(M-1)T_{g+1}^H-T_g^H=M-1+T_{g}^{P}.
\end{equation}
Considering the initial conditions $T_2^P=M+1$ and
$T_2^H=(2M-1)/(M-1)$, we can solve the simultaneous equations, i.e.,
Eqs.~(\ref{eq:MPTP2}) and~(\ref{eq:MPTPQ2}), to obtain
\begin{equation}\label{eq:MPTP3}
T_{g}^P=\left(3M-8+\frac{7M-2}{M^{2}}\right)\left(\frac{M}{M-1}\right)^{g}-2M+3
\end{equation}
and
\begin{equation}\label{eq:MPTPQ3}
T_{g}^H=\left(3-\frac{5M-2}{M^2}\right)\left(\frac{M}{M-1}\right)^{g}-1.
\end{equation}

The obtained expressions for $T_g^P$ and $T_g^H$ are very important,
using which we will determine MFPT $\langle T \rangle_g$. To
facilitate the computation, we use $\Omega_g$ to represent the set
of nodes in $H_g$ and separate them into two subsets: one subset is
$\Omega_{g-1}$ made up of nodes in the original $H_{g-1}$, and the
other subset, denoted by $\bar{\Omega}_{g}$, is the set of nodes of
the $M-1$ copies of $H_{g-1}$. Let $T_i(g)$ denote the trapping time
for a walker originating at node $i$ on the $g$th generation network
to first reach the trap node (hub). Obviously, for all $g \geq 0$,
$T_1(g)=0$. For $g = 1$, it is a trivial case, we have
$T_2(1)=T_3(1)=\cdots=T_M(1)=M-1$. Then, by definition, the MFPT
$\langle T \rangle_g$ can be expressed as
\begin{equation}\label{eq:MFPT05}
 \langle T
\rangle_g=\frac{1}{N_g-1}\sum_{i=2}^{N_g} T_i(g),
\end{equation}
where the sum term $\sum_{i=2}^{N_g} T_i(g)$ can be rewritten as
\begin{eqnarray}\label{eq:MFPT06}
\sum_{i=2}^{N_g} T_i(g)&=&\sum_{i=2}^{N_{g-1}} T_i(g)+\sum_{i \in
\bar{\Omega}_{g}} T_i(g)\nonumber \\
&=&\sum_{i=2}^{N_{g-1}} T_i(g-1)+\sum_{i \in \bar{\Omega}_{g}}
T_i(g),
\end{eqnarray}
which is obvious from the particular construction of the
hierarchical networks. Thus, we have
\begin{equation}\label{eq:MFPT07}
 \langle T
\rangle_g=\frac{N_{g-1}-1}{N_g-1}\langle T
\rangle_{g-1}+\frac{1}{N_g-1}\sum_{i \in \bar{\Omega}_{g}}T_i(g).
\end{equation}
Hence, to obtain an exact solution for $\langle T \rangle_g$, all
that is left is to evaluate the sum in Eq.~(\ref{eq:MFPT07}), with a
goal to first find a recursive relation for $\langle T \rangle_g$.
From Figs.~\ref{network} and~\ref{class}, the sum term on the rhs of
Eq.~(\ref{eq:MFPT07}) can be evaluated as follows:
\begin{widetext}
\begin{equation}\label{eq:MFPT08}
\sum_{i \in
\bar{\Omega}_{g}}T_i(g)=T_{g}^{P}|\mathbb{P}|+\frac{|\mathbb{P}|}{M-1}(T_{g}^{P}+1)+
\sum_{m=1}^{g-2}(M-1)^{g-m-1}\left[(N_{m}-1)\langle T
\rangle_{m}+N_{m} T_{m+1}^{H}+N_{m} T_{g}^{P}\right].
\end{equation}
\end{widetext}
Substituting previously obtained equations for the expressions of
related quantities in Eq.~(\ref{eq:MFPT08}) and combining with
Eq.~(\ref{eq:MFPT06}), we can obtain the following recurrence
relation for $\langle T \rangle_{g}$:
\begin{eqnarray}\label{eq:MFPT09}
&\quad&(N_{g+1}-1)\langle T \rangle_{g+1}-M(N_{g}-1) \langle T
\rangle_{g}\nonumber\\
&=&\frac{2(M-1)^{2}}{M^{2}}
M^{g}\left[(3M-2)\left(\frac{M}{M-1}\right)^{g}-M\right].
\end{eqnarray}
Using the initial condition $\langle T \rangle_{2}=M(M+2)/(M+1)$,
Eq.~(\ref{eq:MFPT09}) is solved inductively to obtain the rigorous
expression for the MFPT:
\begin{eqnarray}\label{eq:MFPT10}
&\quad&\langle
T\rangle_{g}\nonumber\\
&=&\frac{M^{g-3}(M-1)}{M^{g}-1}\bigg[\left(6M^{3}-16M^{2}+14M-4\right)\left(\frac{M}{M-1}\right)^{g}\nonumber\\
&\quad&-(5M^{3}-10M^{2}+4M)-2g\left(M^{2}-M\right)\bigg].
\end{eqnarray}

We have checked our analytic formula against numerical values
obtained according to the fundamental matrix provided by
Eq.~(\ref{MFPT03}). For different parameters $M$ and $g$, the values
obtained from Eq.~(\ref{eq:MFPT10}) completely agree with those
numerical results on the basis of the direct calculation through
Eq.~(\ref{MFPT03}); see Fig.~\ref{Time}. This agreement serves as an
independent test of our theoretical formula. Moreover, we have
performed genuine simulations of the random walk process on the
hierarchical networks. The data from the true process are shown in
Fig.~\ref{Time}, each of which is obtained by averaging over
$10\,000$ realizations. The results of the true simulations are in
excellent agreement with our analytical ones given by
Eq.~(\ref{eq:MFPT10}), and thus provide an important further
evidence in favor of our findings.

We continue to show how to represent MFPT in terms of network order
$N_g$ with the aim to obtain the scaling between these two
quantities. Recalling $N_g=M^g$, we have $g=\log_MN_g$. Hence,
Eq.~(\ref{eq:MFPT10}) can be rewritten as
\begin{eqnarray}\label{eq:MFPT11}
&\quad&\langle
T\rangle_{g}\nonumber\\
&=&\frac{M-1}{M^{3}}\frac{N_g}{N_g-1}\bigg[\left(6M^{3}-16M^{2}+14M-4\right)(N_g)^{1-\frac{\ln (M-1)}{\ln M}}\nonumber\\
&\quad&-(5M^{3}-10M^{2}+4M)-2\left(M^{2}-M\right)\log_MN_g\bigg].
\end{eqnarray}
Thus, for networks with large order, i.e., $N_g\rightarrow \infty$,
\begin{equation}\label{MFPT21}
\langle T \rangle_g \sim (N_g)^{\theta(M)}=(N_g)^{1-\ln (M-1)/\ln
M},
\end{equation}
where the exponent $\theta(M)$ is lower than 1. Clearly, $\theta(M)$
is a decreasing function of $M$: when $M$ grows from 3 to infinite,
$\theta(M)$ descends from $1-\ln 2 /\ln 3$ and approaches to zero,
which means that the efficiency of the trapping process depends on
$M$. The larger the value of $M$, the more efficient the trapping
process. Equation~(\ref{MFPT21}) also implies that in the infinite
network order $N_g$ limit, the MFPT grows algebraically with
increasing order of the networks.

The above obtained scaling of MFPT with order of the hierarchical
scale-free networks is quite different from those scalings for other
media. For instance, on regular lattices with large order $N$, the
asymptotical behavior of MFPT $\langle T \rangle$ is $\langle T
\rangle \sim N^2$, $\langle T \rangle \sim N\ln N$, and $\langle T
\rangle \sim N$ for dimensions $d=1$, $d=2$, and $d=3$,
respectively~\cite{Mo69}. Again for example, on planar Sierpinski
gasket~\cite{KaBa02PRE} and Sierpinski tower~\cite{KaBa02IJBC} in
three Euclidean dimensions, and the $T$ fractal~\cite{Ag08}, the
MFPT $\langle T \rangle$ scales superlinearly with network order;
i.e., it grows as a power-law function of network order with the
exponents being 1.464, 1.293, and 1.631, respectively. Finally, for
the pseudofractal web~\cite{ZhQiZhXiGu09}, the Koch
network~\cite{ZhZhXiChLiGu09}, and the Apollonian
network~\cite{ZhGuXiQiZh09}, they are all scale free, their MFPT
scales linearly or sublinearly with network order, following
separately the asymptotical behaviors $\langle T \rangle \sim N$,
$\langle T \rangle \sim N^{\ln 2 /\ln3}$, and $\langle T \rangle
\sim N^{2- \ln 2/\ln 5}$. Thus, compared with the aforementioned
regular networks, fractals, even scale-free networks, the addressed
hierarchical networks exhibit more efficient configuration for
random walks with a single trap fixed at the node with highest
degree.

The root of the high efficiency of the trapping problem on the
hierarchical scale-free networks lies in their architecture. In this
network family, there are many small densely interconnected
clusters, which combine to form larger but less compact groups
connected by nodes with high degrees (i.e., local hub nodes). The
relatively large groups are further joined to shape even larger and
even less densely interlinked modules. These modules or groups are
combined again at a ``large'' node forming a fine hierarchical
structure that is responsible for the fast diffusion phenomenon,
which can be understood from the following heuristic argument. When
a walker starts off from some node, it will either hit the hub
directly or first get to local hub nodes. These local hubs, although
not connected to the trap node, play the role of bridges linking
different modules together at the local peripheral nodes, through
which the walker may easily find the way to the trap. Thus, the
walker can visit the trap in a short time, disregarding its starting
points.

\section{Conclusions}

In conclusion, we have investigated the classic trapping problem on
a class of hierarchical networks that can bring under a single roof
the scale-free and modular topologies, which are two striking
structural properties observed in various biological and social
networks. Thus, the hierarchical networks can mimic some real-world
natural and social systems to some extent (to what extent it does is
still an open question). Using the method of generating functions,
we derived the recursion relations governing the evolution of the
MFPT for random walks on the networks, with the only trap located at
the hub node. These recursive relations are obtained from the
special construction of the networks, from which we determined
explicitly the solution for the MFPT, which shows that the MFPT
$\langle T \rangle_g $ varies algebraically with network order $N_g$
as $\langle T \rangle_g \sim (N_g)^{\theta(M)}$ with the exponent
$\theta(M)$ much less than 1 that decreases from $1-\ln 2/\ln 3$ to
zero when $M$ increases from 3 to infinite. Thus, in the full range
of $M$, the efficiency of the trapping process on the hierarchical
networks is high. We have also compared the result with those
previously obtained for other media, and found that in marked
contrast to other graphs, the hierarchical networks have more
efficient structure that tends to speed up the diffusion process.
Finally, it deserves to be mentioned that although the hierarchical
networks are efficient for the trapping problem with the trap fixed
on the hub, they might lose this characteristic when the trap is
positioned at a randomly selected node, due to the somewhat
tree-like macro-structure of this kind of
networks~\cite{BaCaPa08,NoRi04a}.

\subsection*{Acknowledgment}

We would like to thank Xing Li for support. This research was
supported by the National Natural Science Foundation of China under
Grants No. 60704044, No. 60873040, and No. 60873070, the National
Basic Research Program of China under Grant No. 2007CB310806,
Shanghai Leading Academic Discipline Project No. B114, and the
Program for New Century Excellent Talents in University of China
(Grants No. NCET-06-0376). S. Y. G. also acknowledges the support by
Fudan's Undergraduate Research Opportunities Program.

\end{document}